\documentclass[twocolumn,a4paper,10pt]{article}
\usepackage{amsmath,amsfonts}
\usepackage{epsfig}
\usepackage{epstopdf}
\usepackage{times}
\usepackage{float}
\usepackage{graphicx}
\usepackage{subcaption}
\usepackage{url}
\usepackage{color}
\makeatletter

\newcounter{numbersec}
\renewcommand{\section}[1]{\par\noindent\stepcounter{numbersec}
\par
\vspace{6pt}
\noindent\textbf{\large   \arabic{numbersec} \hspace*{0.3cm} #1 }
\par
\vspace{2pt}
}
\renewcommand{\subsection}[1]{
\par
\vspace{6pt}
\noindent\textbf{#1}
\par
}
\renewcommand{\subsubsection}[1]{%
\par
\vspace{6pt}
\textbf{#1.}
}

\newcommand{\Abstract}{\par\vspace{6pt}\noindent\textbf{\large Abstract}\par\vspace{2pt}}
\newcommand{\Acknowledgments}{\par\vspace{6pt}\noindent\textbf{\large Acknowledgments }\par\vspace{2pt}}

\newenvironment{References}{
\par\vspace{6pt}\noindent\textbf{\large References}\par\vspace{2pt}
\begin{small}\begin{list}{ }{
\itemsep0mm \parsep0mm\labelsep0mm\leftmargin0mm
}}
{\end{list}\end{small}}

\makeatother

\title{\vspace*{-12mm}
\LARGE \sc \textbf{  
History effects for cambered and symmetric wing profiles
}}
\author{ \Large \bf \textit{ 
A. Tanarro$^*$, R. Vinuesa and P. Schlatter}  \\ \\
\textit{Linn\'e FLOW Centre, KTH Mechanics} \\
\textit{and Swedish e-Science Research Centre (SeRC), SE-100 44 Stockholm, Sweden} \\ \\
\underline{\bf atdr@mech.kth.se}
}
\date{}

\begin{document}
%


%

\maketitle
\thispagestyle{empty}



%
%
\Abstract
The characteristics of complex turbulent boundary layers under adverse pressure gradients are assessed through well-resolved large-eddy simulations (LES) with the spectral-element code Nek5000. Two wing sections are analysed: a NACA0012 at $0^\circ$ angle of attack which presents a mild adverse-pressure gradient (APG) along the chord and a NACA4412 at $5^\circ$ angle of attack with a strong adverse pressure gradient on the suction side, both profiles at $Re_c = 400,000$. The turbulent statistics show that the mild-APG turbulent boundary layer (TBL) of the NACA0012 presents a slight deviation of the velocity fluctuations in the outer region while the strong-APG TBL of the NACA4412 shows significantly larger fluctuations throughout the wall-normal direction with respect to the zero-pressure-gradient (ZPG) TBL. These differences are more substantial in the outer region of the boundary layer. Spectral analyses show that the APG has a significant impact on the largest scales in the boundary layer. Our results indicate that the APG increases the turbulent kinetic energy (TKE) of the TBL, more prominently in the outer layer, and suggest a different mechanism than the one related to high $Re$ in ZPGs.

%
%
\section{Introduction}
Extensive research has been performed in the field of fluid dynamics related to the search of  optimal strategies that allow drag reduction in bluff bodies, such as aircraft and automobiles among others. The direct effect of the drag reduction is the decrease of fuel consumption and, in turn, the emissions of pollutant gases which are and will be subjected to strict regulations. The flows present in industrial applications are typically characterised by very high Reynolds numbers, \textit{Re}. In this type of flows the turbulent boundary layer entails a significant part of the drag generation, and its relevance is supported by numerous studies, experimental and numerical, dedicated to the analysis of turbulent boundary layers (TBLs). Regarding numerical works, these have been possible because of the continuous improvement of computational power, allowing to simulate and analyse most of the relevant characteristics of the flow. Despite the fact that the Reynolds number in the current numerical studies of turbulence is still far from that in the actual application, it is reaching higher values comparable to the range of wind-tunnel experiments. Some examples of this are the analysis of pressure-gradient turbulent boundary layers (PG TBLs) carried out by Vinuesa \textit{et al.} (2017a), who performed a direct numerical simulation (DNS) of the flow around a NACA4412 wing section at  $Re_c = 400,000$ (where $Re_c$ is based on the inflow velocity $U_{\infty}$ and the wing chord $c$) and the large-eddy simulations (LESs) by Sato \textit{et al.} (2016), Fr\`ere (2018) and Vinuesa \textit{et al.} (2018) at $Re_c$ values of at least 1 million.

The aim of the present study is to analyse history effects introduced by adverse pressure gradients (APGs) in turbulent boundary layers. Whereas in some recent studies APG TBLs on flat plates have been analysed (Bobke \textit{et al.}, 2017, Kitsios \textit{et al.}, 2017), in this work we consider the flow around two wing profiles which exhibit different APG distributions along the chord. The wing sections selected for the analysis are the cambered NACA4412 with $5^\circ$ angle of attack and the symmetric NACA0012 with $0^\circ$ incidence, both profiles being simulated at $Re_c = 400,000$. These two flows exhibit different behaviours of the turbulent boundary layers, where their streamwise development (Vinuesa \textit{et al.}, 2017b) and turbulent statistics are significantly affected by the APG distribution present on the suction side, denoted as $(\cdot)_{ss}$.

\section{Numerical Method}

\begin{figure}[h]
\begin{center}
\includegraphics*[width=0.8\linewidth]{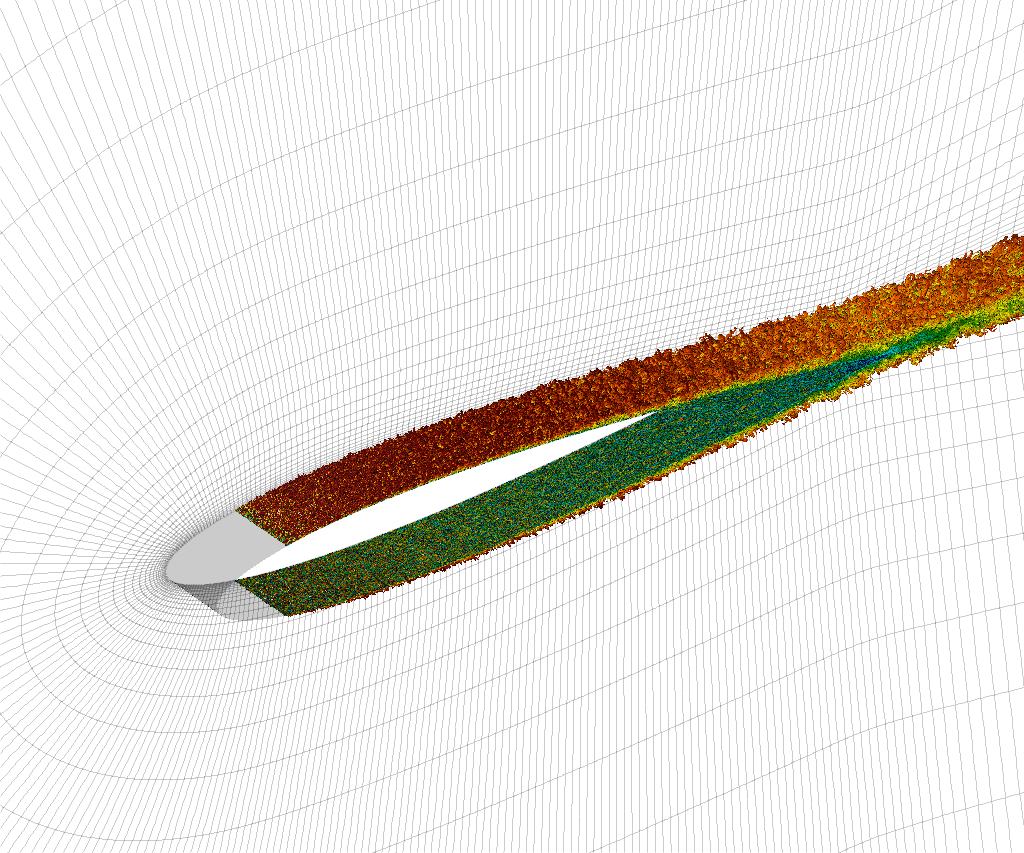}
\caption{\label{fig:graph1} Instantaneous visualisation of the NACA0012 case showing coherent vortical structures identified through the $\lambda_2$ criterion, together with the employed spectral-element mesh (where the grid points within elements are not shown). Structures colored by their streamwise velocity from zero (blue) to the free-stream velocity (red).}
\end{center}
\end{figure}

Successful simulations of turbulent flow are preferably based on the use of high-order numerical methods such that all the scales, or at least all the relevant scales in the flow, are captured by the numerical method. For this reason we perform the analysis of the turbulent boundary layers by means of a well-resolved LES that accurately represents the largest scales of the flow while the smallest scales are modelled. The LES of both airfoils is carried out with the spectral-element code Nek5000 (Fischer \textit{et al.}, 2008). The spatial discretisation is performed by means of Lagrange interpolants of polynomial order $N = 11$ and the spatial resolution in the boundary layer is expressed in wall units: $\Delta x_t^+ = 18.0$, $\Delta y_n^+ = (0.64, 11.0)$ and $\Delta z^+ = 9.0$. Note that $x_t$ and $y_n$ denote the directions tangential and normal to the wing surface, respectively, and $z$ is the spanwise direction. The scaling in wall units is in terms of the friction velocity $u_\tau$ and the viscous length $\ell^{*} = \nu/u_\tau$. The domain under consideration in both cases is a C-mesh (as shown in Figure \ref{fig:graph1} together with an instantaneous flow visualisation) with chordwise length $L_x = 6c$, vertical length $L_y = 4c$ and periodic spanwise length $L_z = 0.1c$. The initial condition is obtained from a RANS simulation from which the values at the boundaries remain as boundary conditions except in the outflow where a stabilised outflow boundary condition is imposed. The domains of the NACA0012 and NACA4412 are discretised using a total of 220,000 and 270,000 spectral elements, respectively, which amount to approximately 380 million and 466 million grid points.

\begin{figure}[h]
\begin{center}
\includegraphics[width=0.8\linewidth]{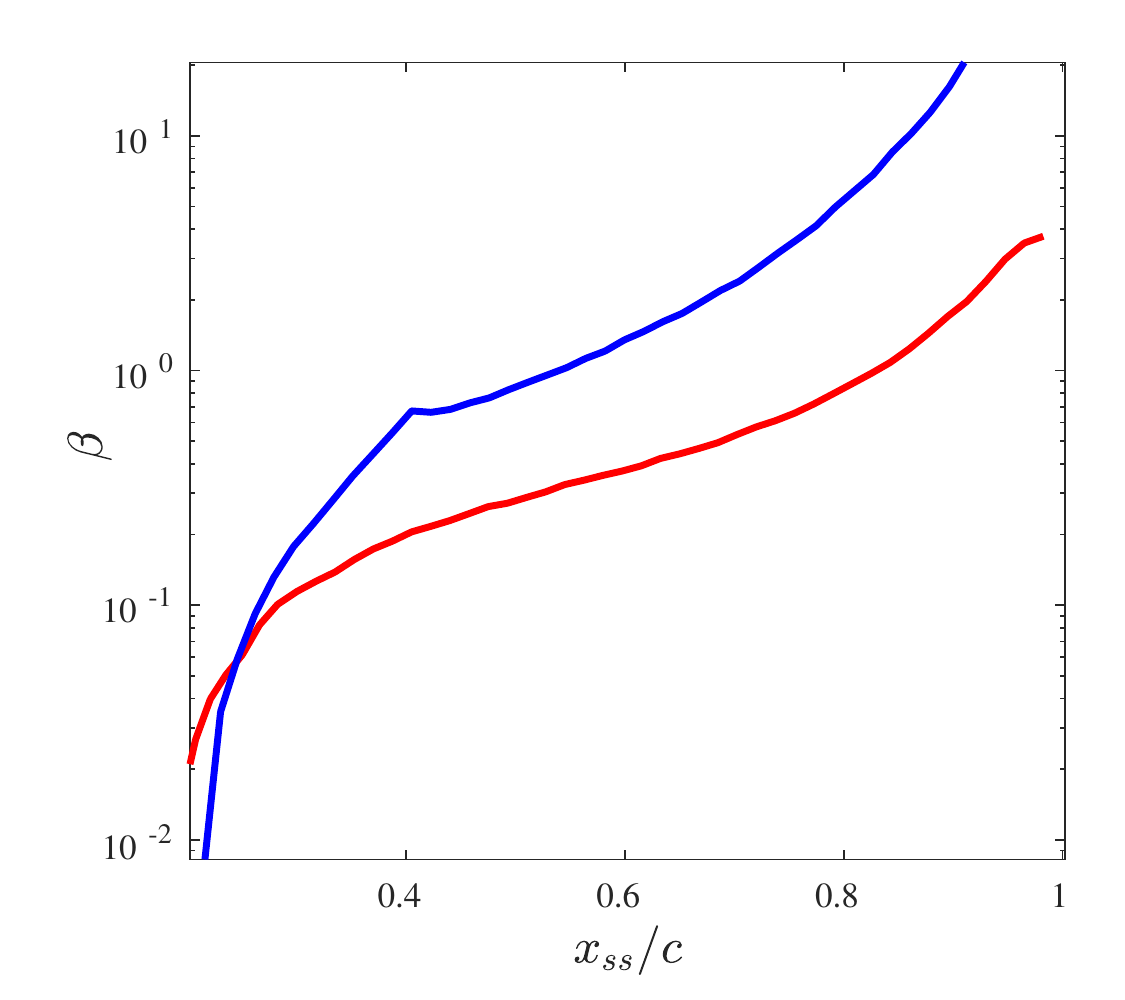}
\caption{\label{fig:graph2} Clauser pressure-gradient parameter $\beta$ along the suction side of the NACA4412 (\textcolor{blue}{---}) and the NACA0012 (\textcolor{red}{---}) wing sections.}
\end{center}
\end{figure}

\section{Turbulence Statistics}
The present results were obtained by averaging for 4.3 flow-over times in the NACA4412 case and 4.0 flow-over times in the NACA0012 (note however that after applying the flow symmetry in the latter, an effective average over 8.0 flow-over times is obtained). The turbulent boundary layer is tripped by means of a weak random volume forcing (see Schlatter and \"Orl\"u, 2012) located on both the pressure and suction sides at $x/c = 0.1$ from the leading edge. Here we focus on the suction side of both wing sections, which exhibit strong APGs with varying magnitudes in the streamwise direction. The turbulent statistics are expressed in the directions tangential ($t$) and normal ($n$) to the wing surface (see Vinuesa \textit{et al.}, 2017a for additional details). The magnitude of the APG is determined in terms of the Clauser pressure-gradient parameter $\beta = \delta^*/\tau_w {\rm d} P_e/{\rm d}x_t$, where $\delta^*$ is the displacement thickness, $\tau_w$ is the wall-shear stress and $P_e$ is the pressure at the boundary-layer edge. As shown in Figure \ref{fig:graph2}, the Clauser pressure-gradient curves show a much stronger APG in the NACA4412 than in the NACA0012, specially as the trailing edge is approached. 


\begin{figure}[h]
\begin{center}
\includegraphics*[width=0.85\linewidth]{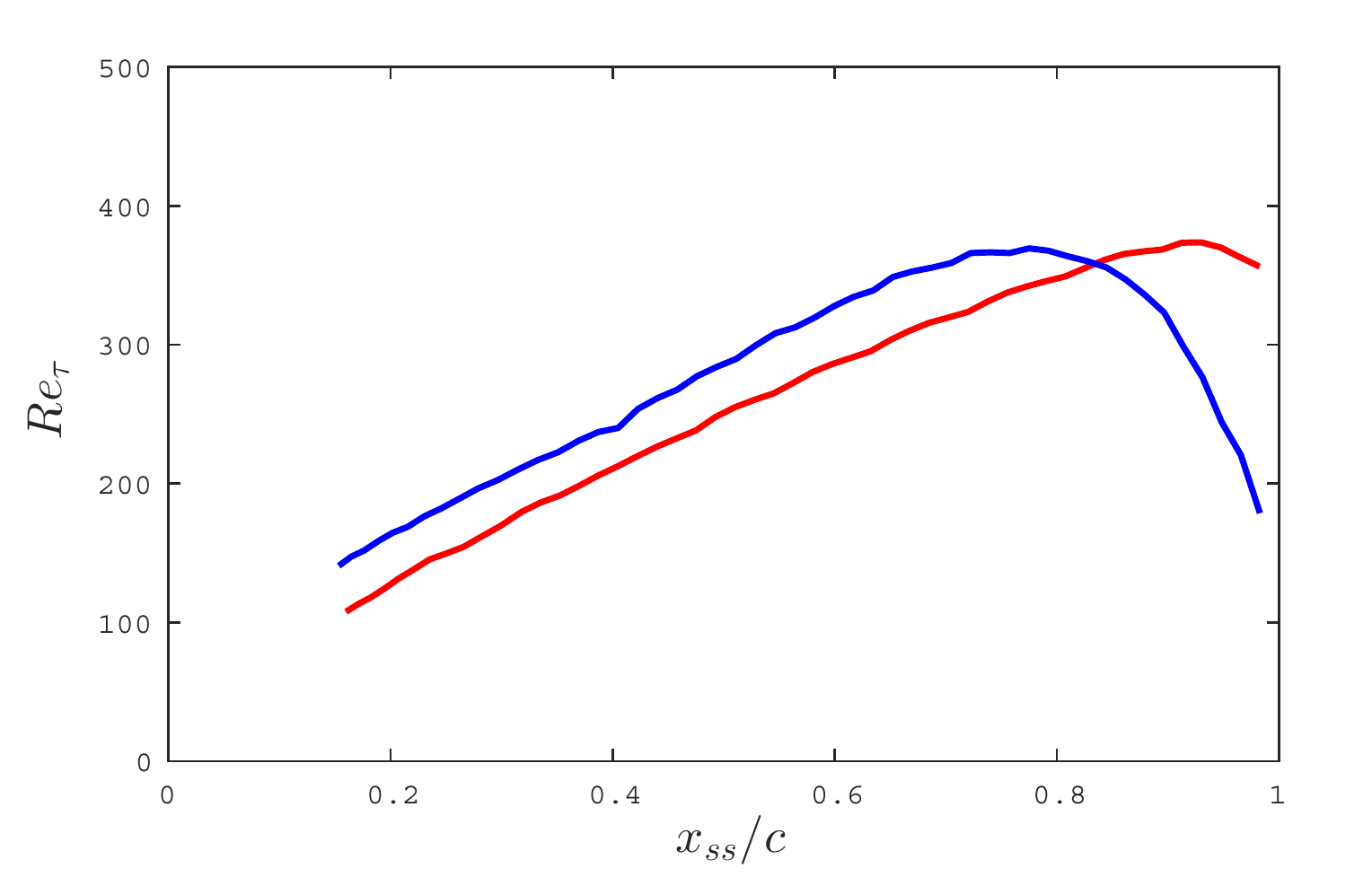}
\caption{\label{fig:graph3} Streamwise evolution of friction Reynolds number. The TBLs developing on the suction side of the NACA4412 (\textcolor{blue}{---}) and the NACA0012 (\textcolor{red}{---}) wing sections are shown.}
\end{center}
\end{figure}

\begin{figure}[h]
\begin{center}
\includegraphics*[width=0.85\linewidth]{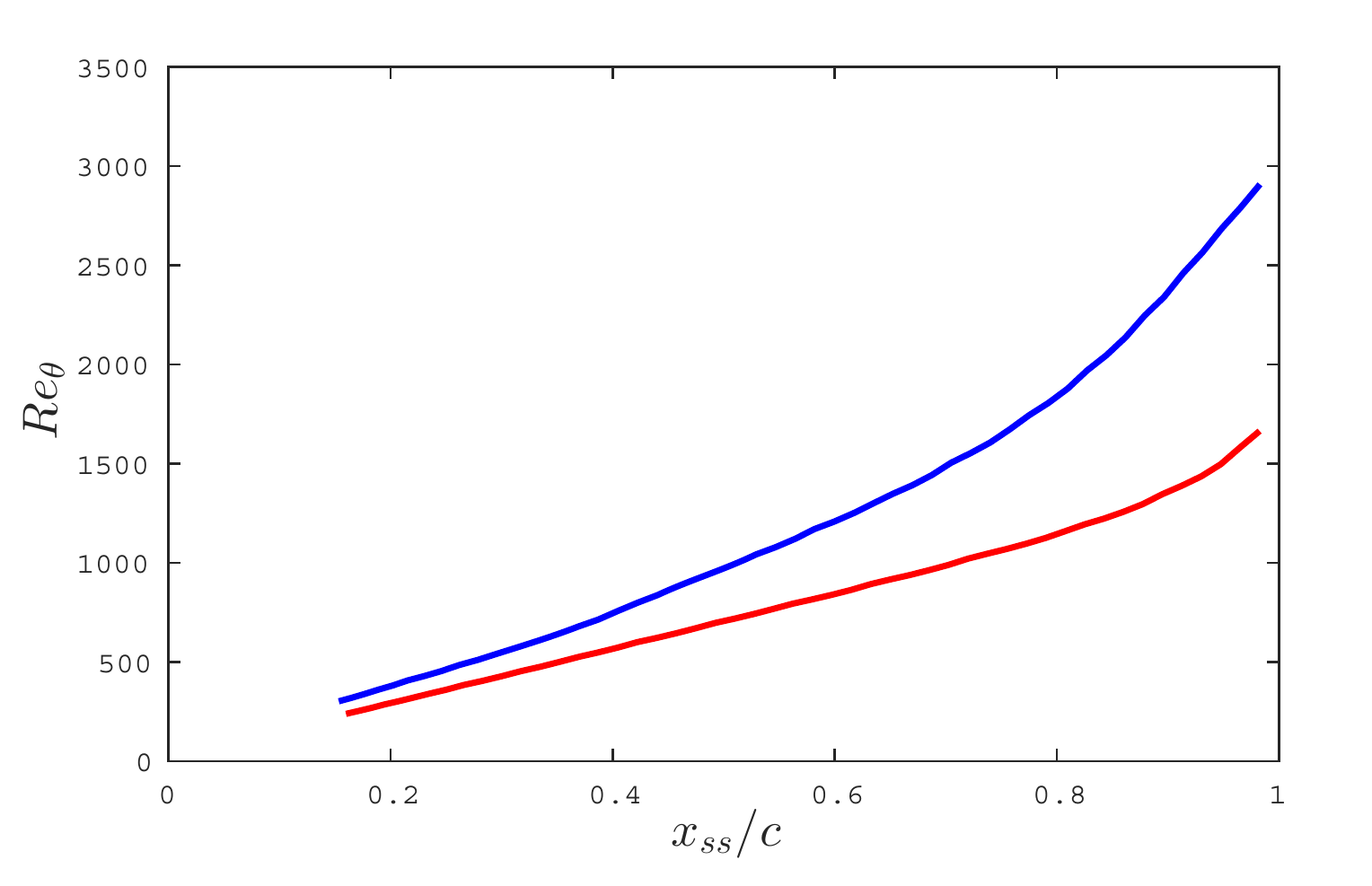}
\caption{\label{fig:graph4} Streamwise evolution of momentum-thickness Reynolds number. The TBLs developing on the suction side of the NACA4412 (\textcolor{blue}{---}) and the NACA0012 (\textcolor{red}{---}) wing sections are shown.}
\end{center}
\end{figure}

The friction Reynolds number (defined in terms of $u_\tau$ and the $99\%$ boundary-layer thickness $\delta_{99}$) is shown in Figure \ref{fig:graph3}. Here we employ the method by Vinuesa \textit{et al.} (2016) to determine the boundary-layer thickness. The $Re_\tau$ curve of the NACA4412 reaches a maximum value of $Re_\tau = 369$ at $x_{ss}/c \simeq 0.8$ before showing a sudden drop because of the very strong APG in that region, while the NACA0012 only shows a subtle decrease beyond $x_{ss}/c \simeq 0.93$ where $Re_\tau = 372$. In the case of the momentum-thickness-based Reynolds number $Re_{\theta}$ (Figure \ref{fig:graph4}) one can observe the difference between the two profiles in which both the rate of growth and the local values are significantly higher in the NACA4412 wing section, having as maximum $Re_\theta = 2916$ while the NACA0012 maximum $Re_\theta$ only reaches $1669$. Note that beyond $x_{ss}/c \simeq 0.8$, the cambered NACA4412 exhibits values of $\beta$ an order of magnitude larger than those in the symmetric NACA0012.

\begin{figure}[h]
\begin{center}
\begin{subfigure}[t]{1\linewidth}
\includegraphics*[width=0.95\linewidth]{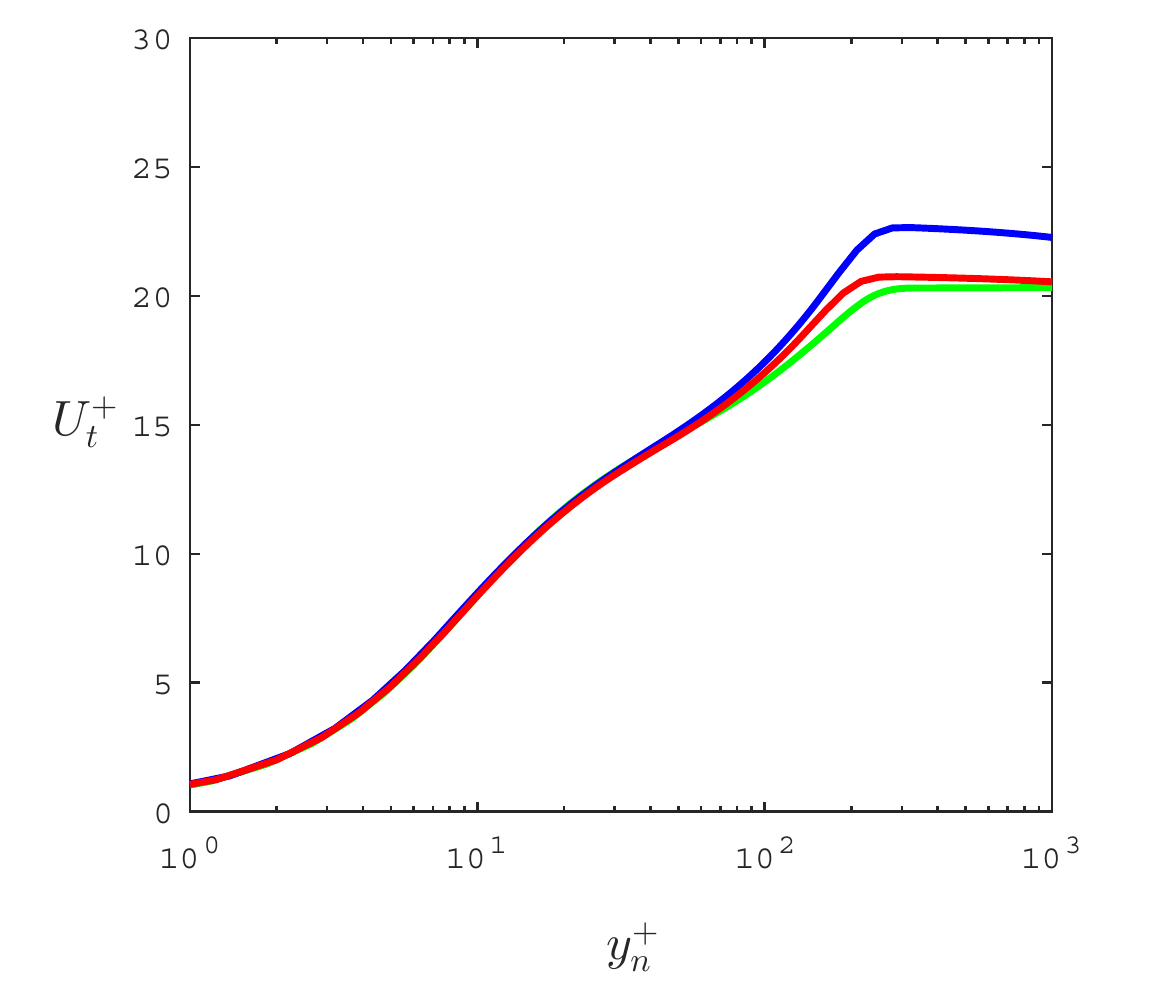}
\caption{\label{fig:graph5} $x_{ss}/c = 0.4$}
\end{subfigure}

\begin{subfigure}[t]{1\linewidth}
\includegraphics*[width=0.95\linewidth]{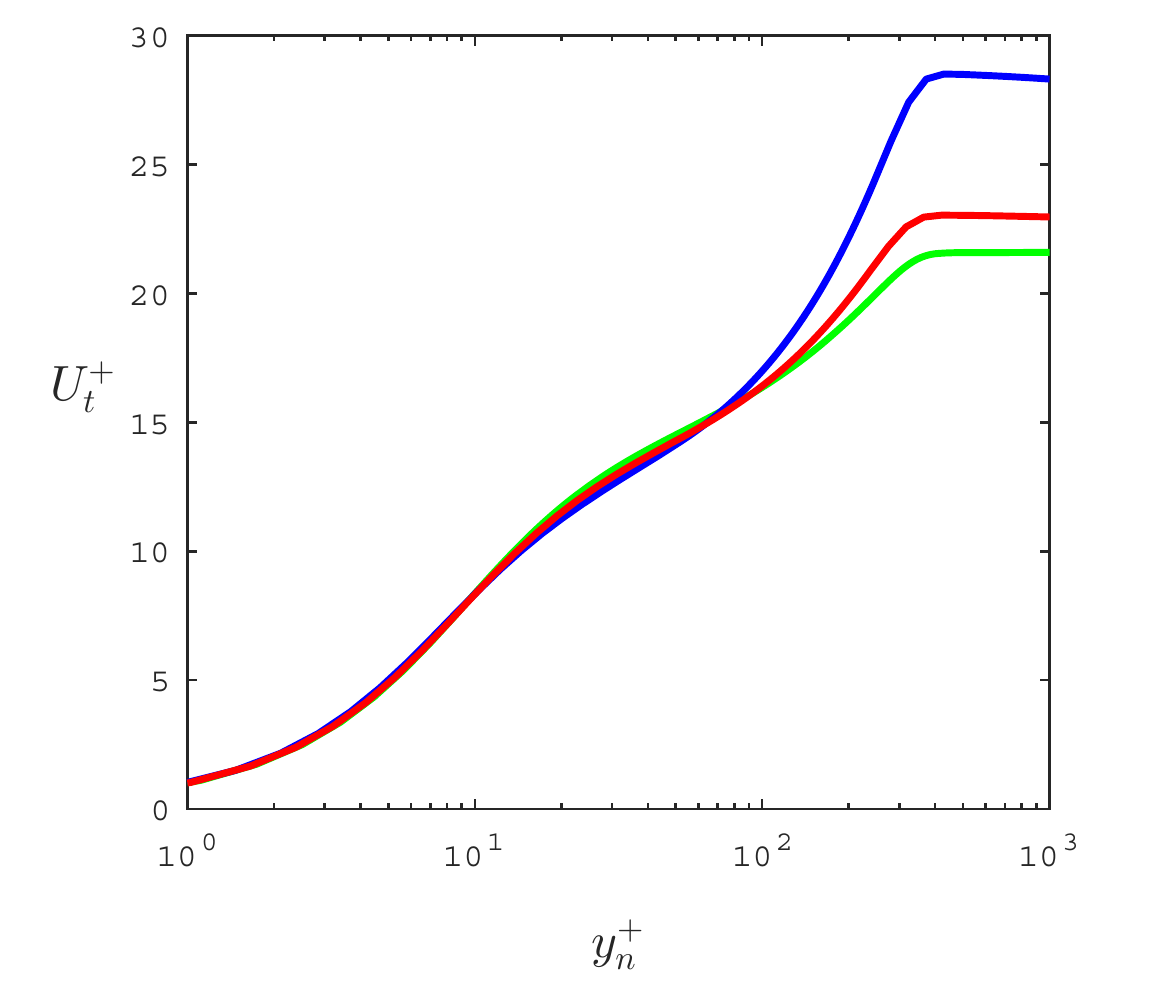}
\caption{\label{fig:graph6} $x_{ss}/c = 0.75$}
\end{subfigure}
\caption{Inner-scaled mean velocity profile along the wall-normal direction in the following cases:  ZPG (\textcolor{green}{---}), NACA4412 (\textcolor{blue}{---}), NACA0012 (\textcolor{red}{---}). Note that the ZPG case was chosen to approximately match the corresponding $Re_\tau$ values of the wing sections.}
\end{center}
\end{figure}

On the other hand, Figures \ref{fig:graph5} and \ref{fig:graph6} show the effect of the APG on the development of the boundary layer along the chord of the wing section compared to the results of a zero-pressure-gradient (ZPG) TBL LES performed by Eitel-Amor \textit{et al.} (2014). It is clear that, at $x_{ss}/c = 0.4$ (Figure \ref{fig:graph5}), near the wall and in most of the incipient overlap region both boundary layers show an almost identical inner-scaled mean velocity profile, while the outer region of the NACA4412 (for which $Re_\tau = 253$ and $\beta = 0.66$) shows a higher inner-scaled mean velocity than the NACA0012 ($Re_\tau = 211$ and $\beta = 0.20$). This effect is more prominent at $x_{ss}/c = 0.75$ (Figure \ref{fig:graph6}), where the stronger-APG effects ($\beta = 4.11$ and $Re_\tau = 369$ for the NACA4412, $\beta = 0.66$ and $Re_\tau = 338$ for the NACA0012) also lead to reduced values in the buffer region (Spalart and Watmuff, 1993). As discussed by Harun \textit{et al.} (2013), the APG increases the TKE in the outer region of the TBL, generating high-energy turbulent structures. The relation between this effect and the inner-scaled mean velocity profile was reported by Bobke \textit{et al.} (2017), who stated that the development of large-scale energetic turbulent structures in the boundary layer differs depending on the APG evolution. 

\begin{figure}[h]
\begin{center}
\begin{subfigure}[t]{1\linewidth}
\includegraphics*[width=0.95\linewidth]{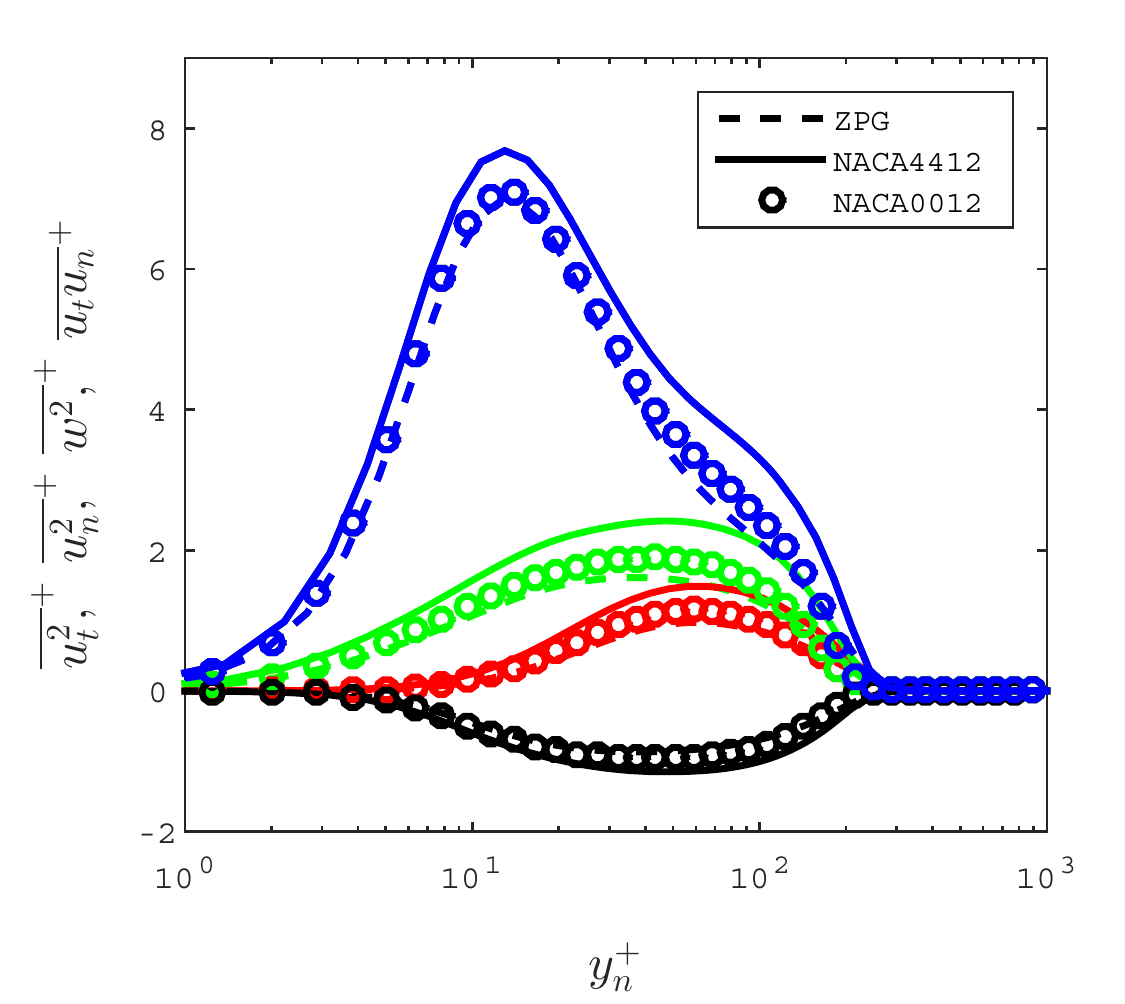}
\caption{\label{fig:graph7} $x_{ss}/c = 0.4$}
\end{subfigure}

\begin{subfigure}[t]{1\linewidth}
\includegraphics*[width=0.95\linewidth]{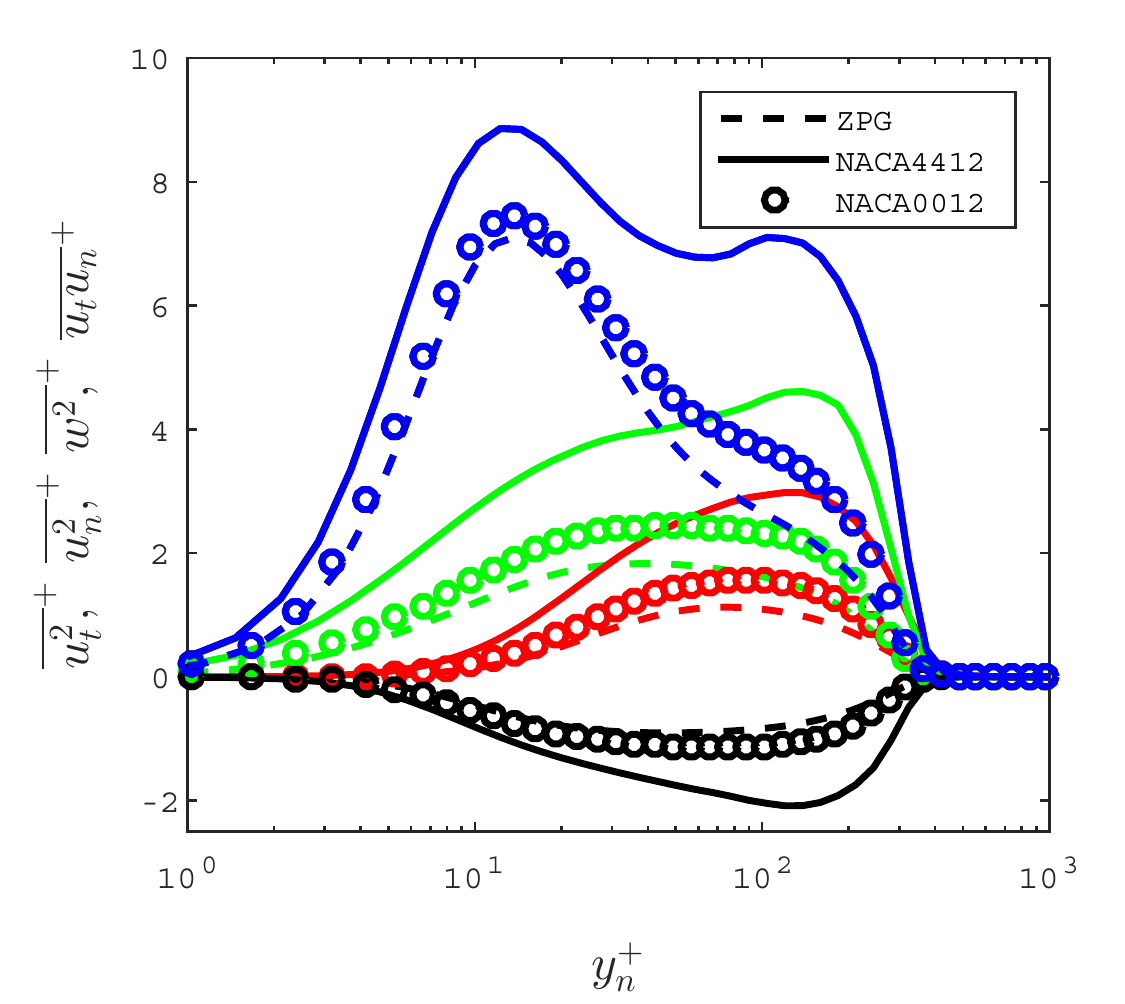}
\caption{\label{fig:graph8} $x_{ss}/c = 0.75$}
\end{subfigure}
\caption{Selected inner-scaled Reynolds stresses: tangential (\textcolor{blue}{---}), normal (\textcolor{red}{---}), spanwise (\textcolor{green}{---}) velocity fluctuations and Reynolds-shear stress (\textcolor{black}{---}).}
\end{center}
\end{figure}

The increment in TKE in the outer region of the two TBLs is analysed through several components of the Reynolds-stress tensor for both wing profiles at $x_{ss}/c = 0.4$ and $0.75$. Figure \ref{fig:graph7} shows the significant difference in the turbulence features as the TBLs develop so that at $x_{ss}/c = 0.4$ we already obtain slightly greater velocity fluctuations when there is a moderate APG. On the other hand, Figure \ref{fig:graph8} shows that at $x_{ss}/c = 0.75$ ({\it i.e.} different $Re_\theta$ and $\beta$ values and very similar values of $Re_\tau \simeq 350$ in both wing cases) noticeable discrepancies among the wings are observed. It is clear how strong APGs increase the TKE of the boundary layer so that the NACA4412 case exhibits values roughly twice as large in the outer region of all the shown Reynolds-stress profiles than the NACA0012 airfoil. The presence of an outer peak in the profile of the tangential velocity fluctuations when imposing a strong APG with $\beta \simeq 4.11$ is also remarkable (Figure \ref{fig:graph8}) as shown by Vinuesa \textit{et al.} (2017a).

\section{Spectral Analysis}
Through the analysis of the statistics of the turbulent boundary layers we have observed a significant effect of the APG on the outer region. The objective now is to identify how the energy is distributed among the scales and which is the effect of the APG in this energy distribution. For that purpose two approaches are considered for the spectral analysis: first, the computation of the inner-scaled pre-multiplied spectra of the tangential velocity fluctuations, $k_z\phi_{u_tu_t}^+$, as a function of the inner-scaled normal distance to the wall $y_n^+$ and the inner-scaled spanwise wavelength $\lambda_z^+$; second, the two-dimensional inner-scaled pre-multiplied spectra, $k_zk_t\phi_{u_tu_t}^+$, as a function of the inner-scaled spanwise wavelength and the inner-scaled temporal period $\lambda_t^+$. The computation of the power-spectral density in terms of the spanwise wavelength is performed by means of Fast Fourier Transform (\textit{i.e.} FFT) due to the periodic boundary conditions imposed in this direction. On the other hand, the same method cannot directly be applied to the temporal signal as it is not periodic nor reaches infinity, therefore the spectra in terms of the temporal period is computed by means of the Welch's method, \textit{i.e.} an overlapping window method. Note that we will be comparing with the ZPG-TBL LES performed by Eitel-Amor \textit{et al.} (2014) at $Re_\tau = 350$ which is similar to the $Re_\tau$ of both wing sections at $x_{ss}/c = 0.75$ but larger than the wing sections at $x_{ss}/c = 0.4$.

\begin{figure}[h]
\begin{center}
\begin{subfigure}[t]{1\linewidth}
\includegraphics*[width=0.95\linewidth]{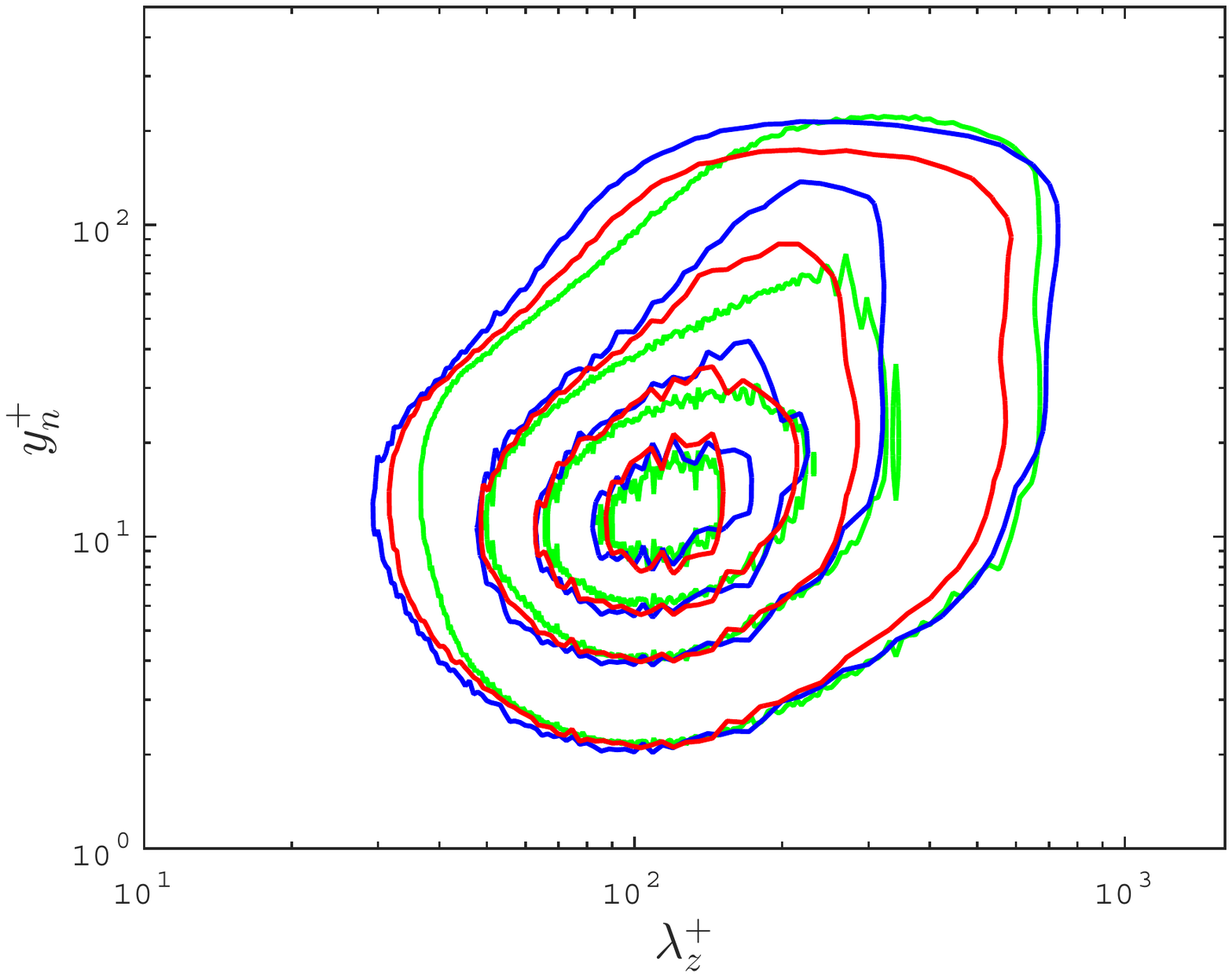}
\caption{\label{fig:graph9} $x_{ss}/c = 0.4$}
\end{subfigure}\vspace{3mm}

\begin{subfigure}[t]{1\linewidth}
\includegraphics*[width=0.95\linewidth]{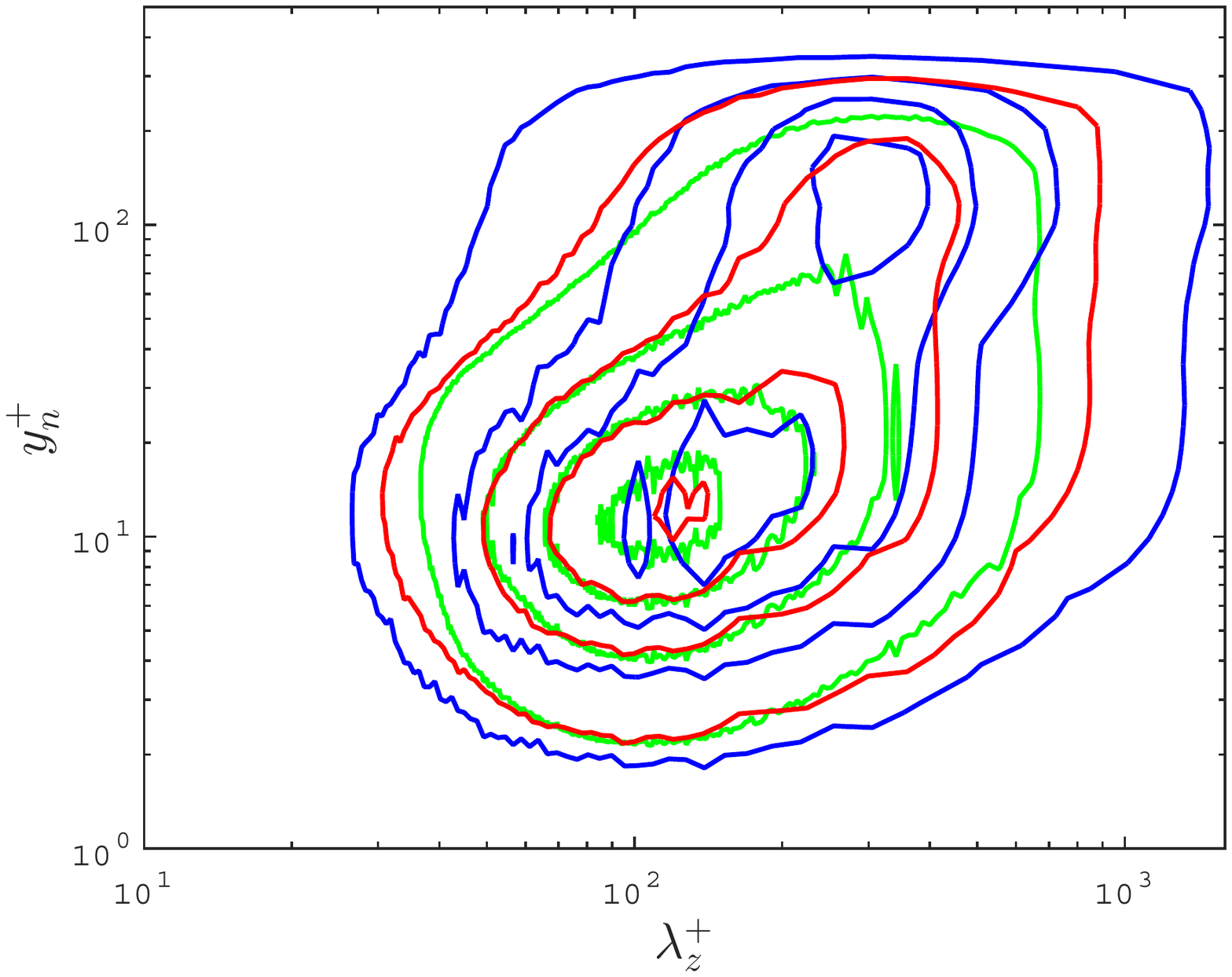}
\caption{\label{fig:graph10} $x_{ss}/c = 0.75$}
\end{subfigure}
\caption{Pre-multiplied spanwise spectra of the tangential velocity fluctuations for the NACA4412 (\textcolor{blue}{---}), NACA0012 (\textcolor{red}{---}) and ZPG at $Re_\theta = 880$ (\textcolor{green}{---}). The contours range from $k_z\phi_{u_tu_t}^+ = 0.5$ to $3.8$ with a constant increment of 1.1 units.}
\end{center}
\end{figure}

First, we consider the pre-multiplied spectra as a function of the spanwise wavelength. Figure \ref{fig:graph9} shows the spectra at $40\%$ of the chord length. In this case, we can observe similar power-spectral density distributions for both wings. This is highly related to the low accumulated APG up to this position which seems not to affect considerably neither the turbulence statistics nor its spectra, although some discrepancies between both cases are observed. For example, the inner peak is very similar in the three cases, however the largest spanwise wavelengths show higher energy levels in the ZPG and the NACA4412. This behaviour is the result of the higher $Re_\theta$ in the ZPG and NACA4412 cases with respect to the NACA0012, in which the APG magnitude is very low.

On the other hand, Figure \ref{fig:graph10} shows more relevant results as it compares the effect of the strong APG (\textit{i.e.} $\beta = 4.11$ for the NACA4412) with the effect of the mild APG (\textit{i.e.} $\beta = 0.66$ for the NACA0012), at $75\%$ of the chord length. The NACA0012 exhibits a slightly higher, but similar in shape, power-spectral density distribution compared to that at $x_{ss}/c = 0.4$ while the power-spectral density distribution in the NACA4412 features more significant differences in the energy levels. The NACA4412 shows higher-energy levels for both small and large scales when compared with the NACA0012 and ZPG. This increment in TKE takes place both in the near-wall region and in the outer region, being similar for the largest spanwise wavelengths but more significant in the outer region than in the near-wall region for the smallest scales. It is important to notice the emergence of the spectral outer peak in the spectra of the NACA4412. 
According to Hutchins \& Marusic (2007) the spectral outer peak is always present but at low Reynolds numbers, the position of both peaks coincide (there is no separation of scales) and they overlap. Their experiment is performed at $Re_\theta = 7100$, which is around 5 times higher than in our simulations and still the peak achieved for the NACA4412 is more prominent. Comparing now the ZPG with the NACA0012, the results lead to a different behaviour than at $x_{ss}/c = 0.4$. Since both cases have a similar $Re_\tau$, the higher TKE in the outer region of the NACA0012 is a result of the APG. 

\begin{figure}[h]
\begin{center}
\begin{subfigure}[t]{1\linewidth}
\includegraphics*[width=0.95\linewidth]{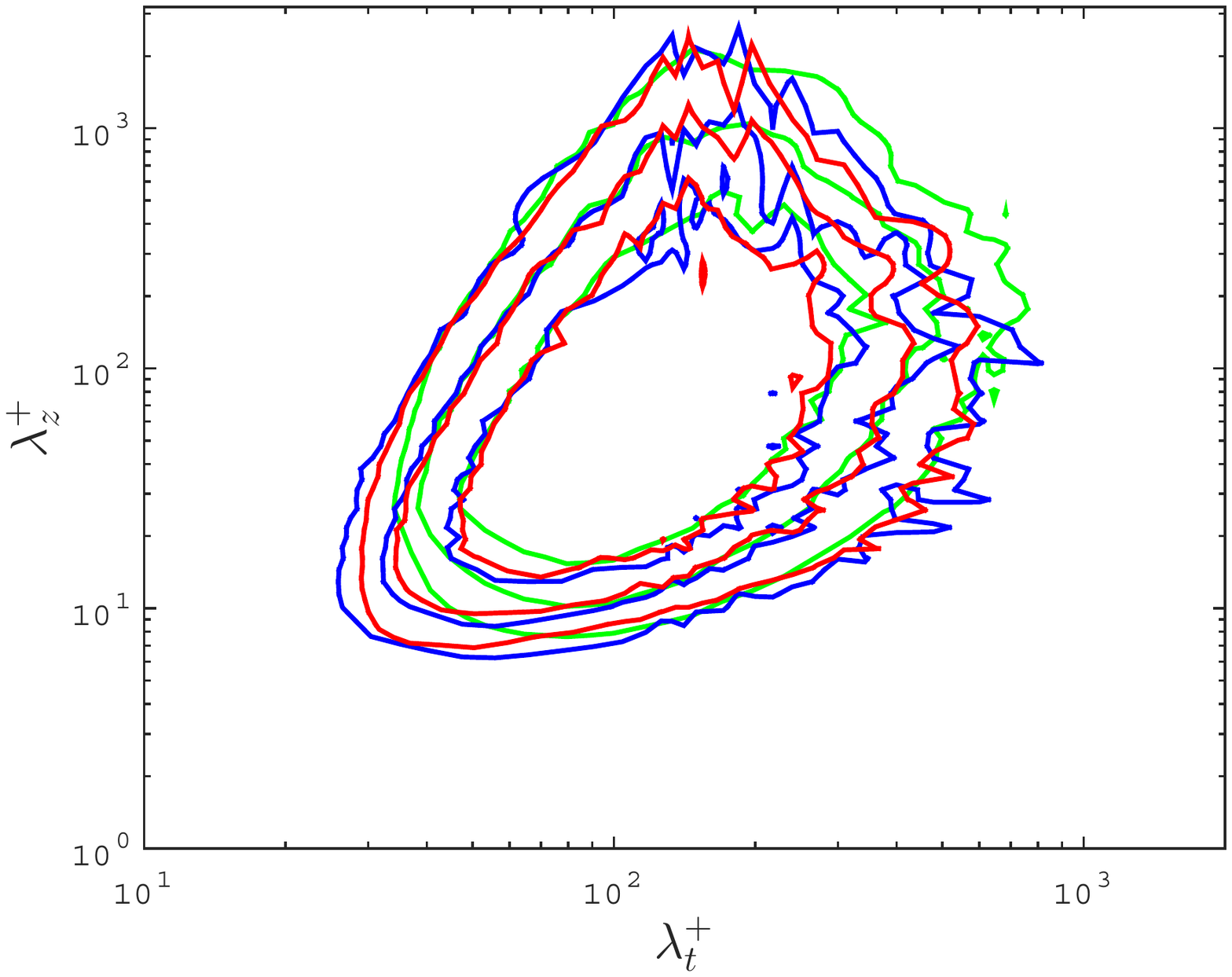}
\caption{\label{fig:graph11} $x_{ss}/c = 0.4$}
\end{subfigure}\vspace{2mm}

\begin{subfigure}[t]{1\linewidth}
\includegraphics*[width=0.95\linewidth]{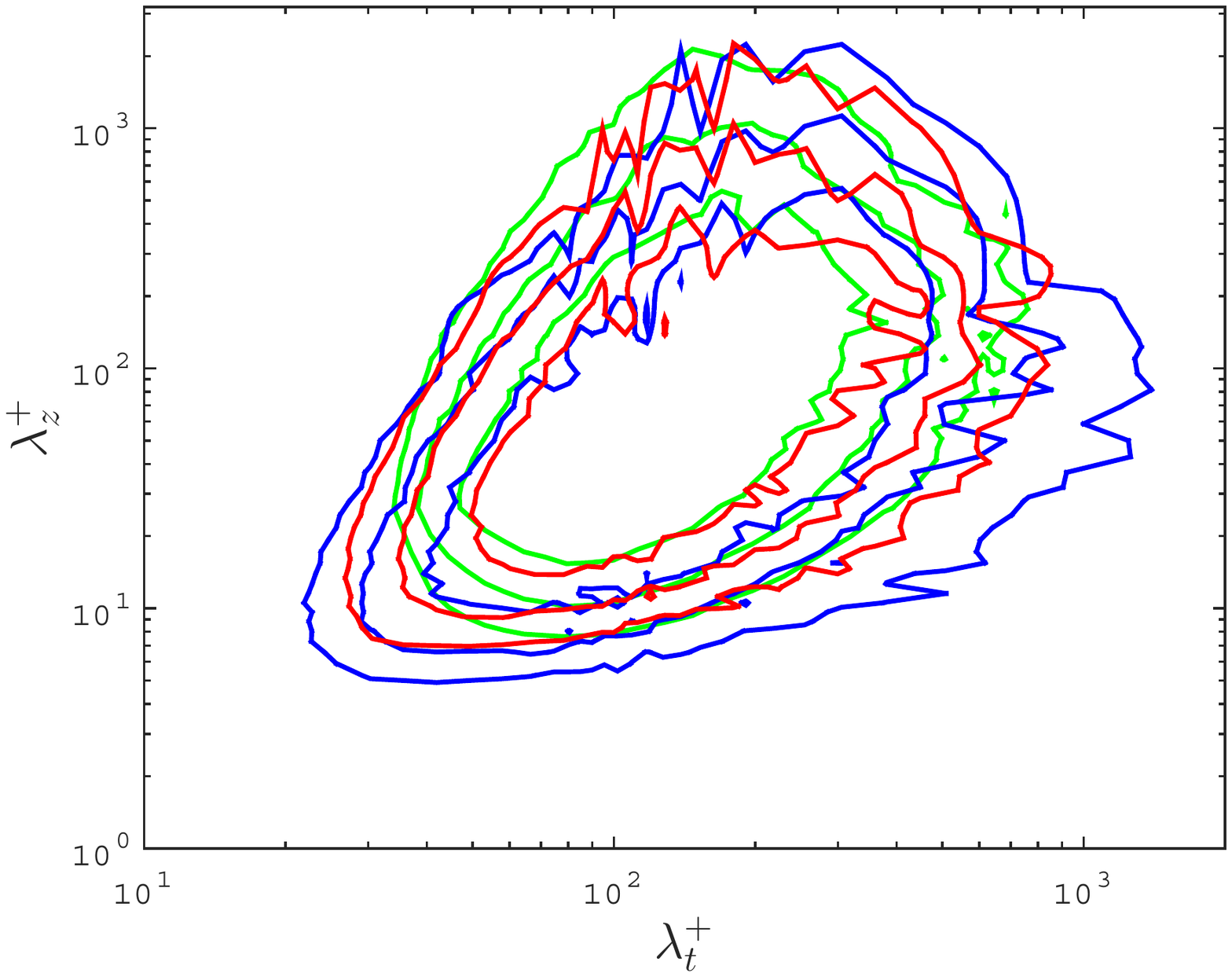}
\caption{\label{fig:graph12} $x_{ss}/c = 0.75$}
\end{subfigure}
\caption{Pre-multiplied two-dimensional spectra of the tangential velocity fluctuations for the NACA4412 (\textcolor{blue}{---}), NACA0012 (\textcolor{red}{---}) and ZPG (\textcolor{green}{---}) at $y_n^+ = 16$. The contours represent $k_zk_t\phi_{u_tu_t}^+ = 0.05$, $0.1$ and $0.2$.}
\end{center}
\end{figure}

Finally, we will study the two-dimensional pre-multiplied spectra of the tangential velocity fluctuations. In this case, we will be computing the spectra as a function of the spanwise wavelength and temporal period at two positions from the wall: in the near-wall region ($y_n^+ = 16$) and in the outer region ($y_n/\delta_{99}=0.2$). Starting with the analysis of the inner region, Figure \ref{fig:graph11} shows similar power-spectral densities for both wing sections at a location in which the accumulated pressure gradient is still considerably low, such that the effect on the TBLs is small. However, comparing with the ZPG the main difference appears when both scales are small. It can be seen that for the wing sections the smallest scales have higher energy levels than those from the ZPG. 
Due to the low APG that both wing sections exhibit at that position and the similarity between both power-spectral densities, a possible explanation is that this effect could come from the surface curvature.

Figure \ref{fig:graph12} shows that this effect is amplified for strong APGs (\textit{i.e.} NACA4412) while it is barely noticeable for the NACA0012. This suggests that the APG also affects the smallest scales, an effect that does not take place when increasing $Re$, and, therefore, that the turbulent production mechanism is different when increasing $Re$ and when increasing the APG, a conclusion similar to the one obtained by Vinuesa \textit{et al.} (2018).  Moreover, there is an increment in the energy levels of the temporal largest scales and an increment in the energy levels of the spanwise smallest scales for the NACA4412.

\begin{figure}[h]
\begin{center}
\begin{subfigure}[t]{1\linewidth}
\includegraphics*[width=0.95\linewidth]{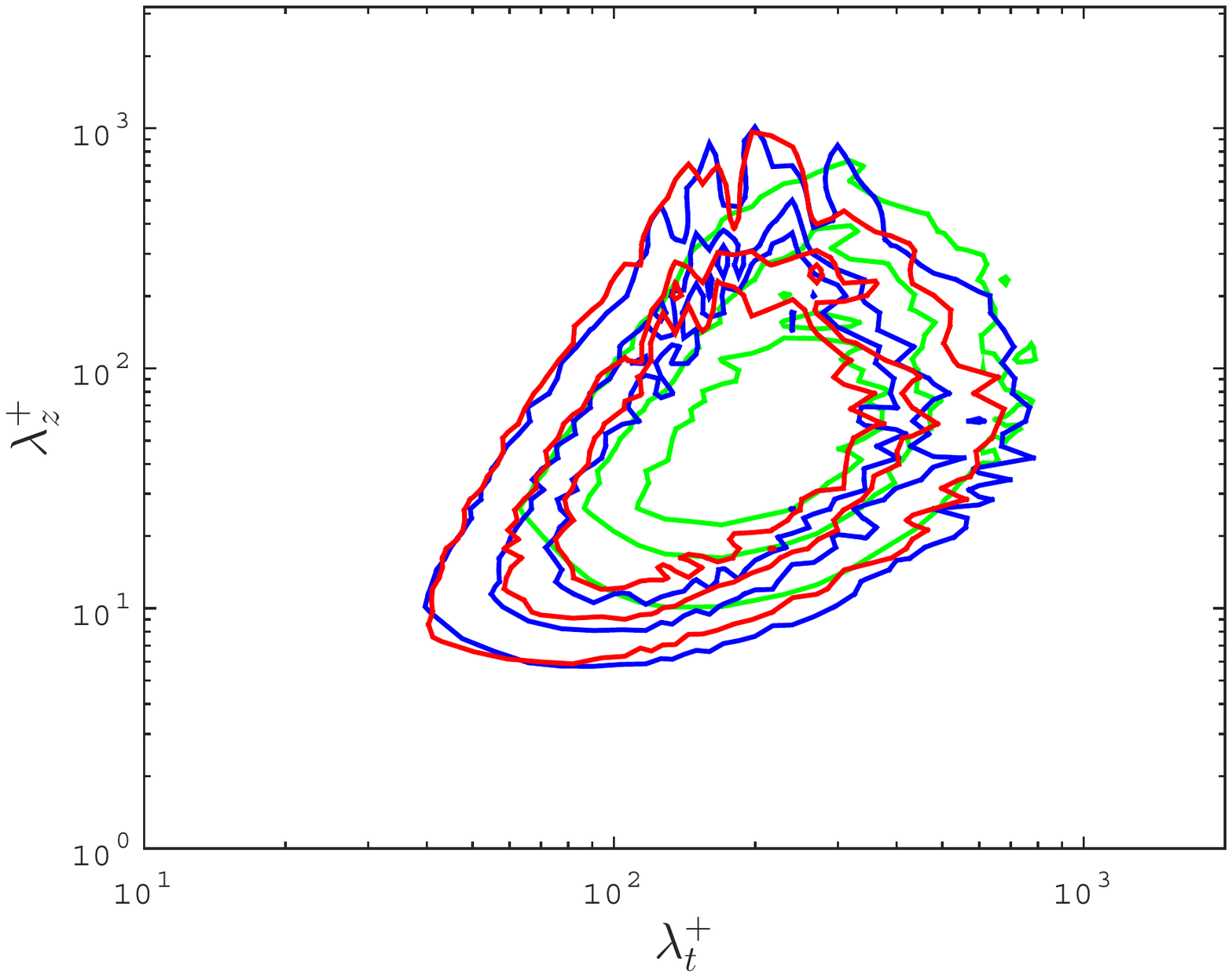}
\caption{\label{fig:graph13} $x_{ss}/c = 0.4$}
\end{subfigure}\vspace{2mm}

\begin{subfigure}[t]{1\linewidth}
\includegraphics*[width=0.95\linewidth]{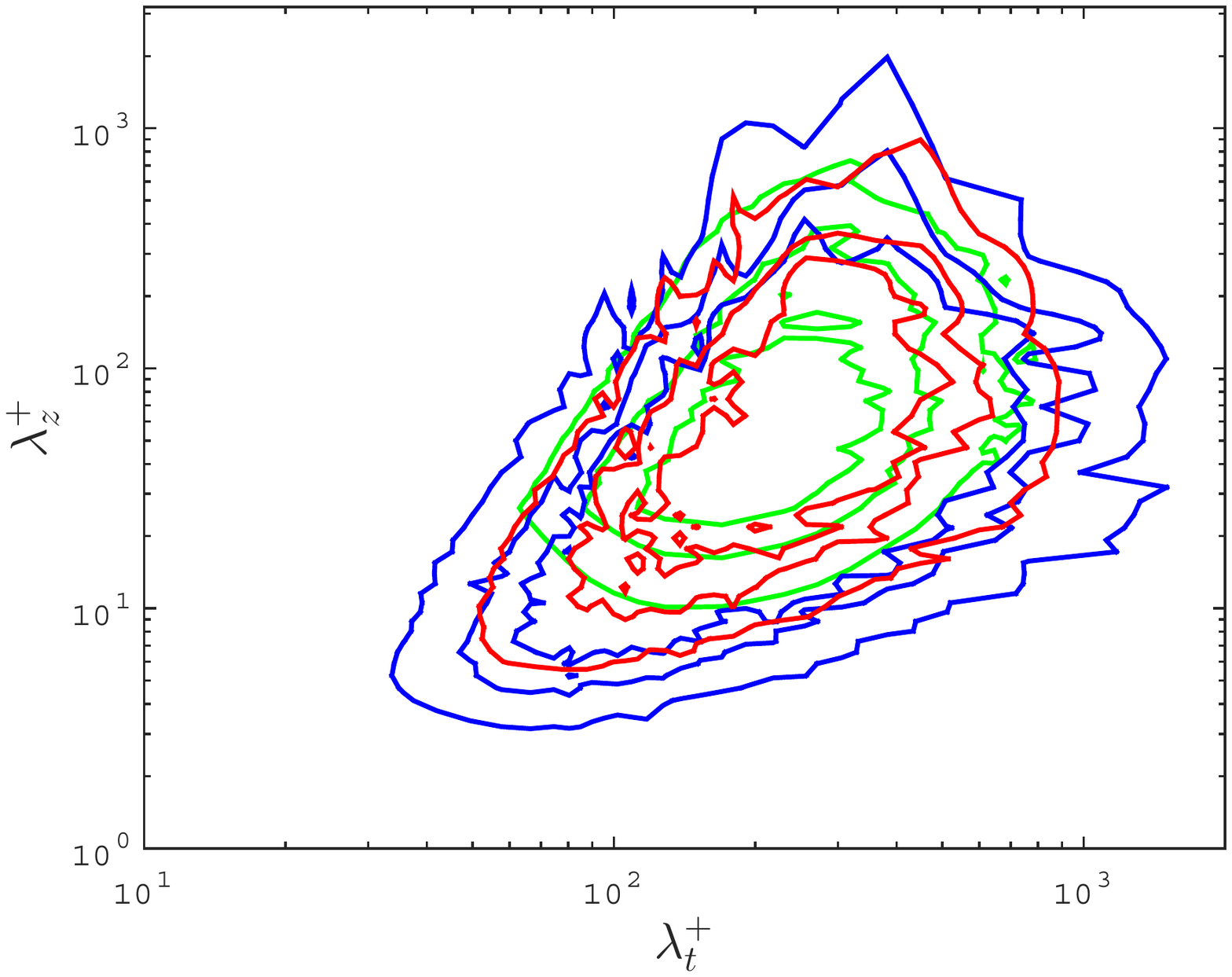}
\caption{\label{fig:graph14} $x_{ss}/c = 0.75$}
\end{subfigure}
\caption{Pre-multiplied two-dimensional spectra of the tangential velocity fluctuations for the NACA4412 (\textcolor{blue}{---}), NACA0012 (\textcolor{red}{---}) and ZPG (\textcolor{green}{---}) at $y_n/\delta_{99} = 0.2$. The contours range from $k_zk_t\phi_{u_tu_t}^+ = 0.05$ to $0.15$ with a constant increment of 0.1 units.}
\label{fig:graph15}
\end{center}
\end{figure}

Focusing now on the outer region (Figure \ref{fig:graph15}), the first aspect to be noticed is the larger energy levels of the wing-section boundary layers compared to the ZPG TBL. Again, as in the inner region, the main difference between the wing sections and the ZPG is in the region where both scales are very low. In this case it can be observed that the energy of all scales is increased, just as shown in Figure \ref{fig:graph10}. However, the different scales of the NACA0012 increase their kinetic energy except large spanwise wavelengths, which have lower energy levels than those at $x_{ss}/c = 0.4$. 

\section{Conclusion}
The present study analyses the effect of APGs on the development of TBLs. The cases under study are the flow over a NACA4412 at $5^\circ$ angle of attack and the flow over a NACA0012 at $0^\circ$, both at $Re_c = 400,000$, using the results of a ZPG LES as reference to study the effect of the APG. The turbulence statistics on the suction side are computed and clearly reflect the impact of the APG. In the case of the mild-APG (\textit{i.e.} NACA0012) the difference with the ZPG BL is smaller, and it is more noticeable in the outer region through increased tangential velocity fluctuations. Furthermore, the strongest APG case shows a large increment in TKE of the boundary layer, mainly in the outer region where an outer peak emerges. The analysis is extended through the computation of the pre-multiplied spanwise and two-dimensional spectra of the tangential velocity fluctuations. The main conclusions extracted from this analysis are the increment in TKE due to the adverse pressure gradient with respect to the ZPG. In addition, an important result is the increment in spectral density in the near-wall region by the APG, an effect that is not observed by increasing $Re$, which suggests that the turbulent production mechanisms due to $Re$ and APG are different.
%

\Acknowledgments
The simulations were performed on resources provided by the Swedish National Infrastructure for Computing (SNIC) at the Center for Parallel Computers (PDC), in KTH, Stockholm. This research is funded by the Swedish Research Council (VR) and the Knut and Alice Wallenberg Foundation.
\begin{References}
\item Bobke, A., Vinuesa, R., \"Orl\"u, R., \& Schlatter, P. (2017). History effects and near equilibrium in adverse-pressure-gradient turbulent boundary layers. \textit{J. Fluid Mech.}, 820, 667--692.

\item Eitel-Amor, G., \"Orl\"u, R., \& Schlatter, P. (2014). Simulation and validation of a spatially evolving turbulent boundary layer up to $Re_\theta= 8300$. \textit{Int. J. Heat Fluid Flow}, 47, 57--69.

\item Fischer, P. F., Lottes, J. W., \& Kerkemeier, S. G. (2008) NEK5000: {O}pen {S}ource spectral element {CFD} solver. Available at: \url{http://nek5000.mcs.anl.gov}

\item Fr\`ere, A. (2018). Towards wall-modelled Large-Eddy Simulations of high Reynolds number airfoils using a discontinuous Galerkin method, PhD thesis, Universit\'e catholique de Louvain, Louvain.

\item Harun, Z., Monty, J. P., Mathis, R., \& Marusic, I. (2013). Pressure gradient effects on the large-scale structure of turbulent boundary layers. \textit{J. Fluid Mech.}, 715, 477--498.

\item Hutchins, N., \& Marusic, I. (2007). Large-scale influences in near-wall turbulence. \textit{Philos. Trans. Royal Soc. A}, 365(1852), 647--664.

\item Kitsios, V., Sekimoto, A., Atkinson, C., Sillero, J. A., Borrell, G., Gungor, A. G., Jimenez, J., \& Soria, J. (2017). Direct numerical simulation of a self-similar adverse pressure gradient turbulent boundary layer at the verge of separation. \textit{J. Fluid Mech.}, 829, 392--419.

\item Sato, M., Asada, K., Nonomura, T., Kawai, S., \& Fujii, K. (2016). Large-Eddy Simulation of NACA 0015 Airfoil Flow at Reynolds Number of 1.6× 10 6. \textit{AIAA J.}, 55(2), 673--679.


\item Schlatter, P., \& \"Orl\"u, R. (2012). Turbulent boundary layers at moderate Reynolds numbers: inflow length and tripping effects. \textit{J. Fluid Mech.}, 710, 5--34.

\item Spalart, P.R. \& Watmuff, J.H. (1993) Experimental and numerical study of a turbulent boundary layer with pressure gradients. \textit{J. Fluid Mech.} 249, 337--371.

\item Vinuesa, R., Bobke, A., \"Orl\"u, R., \& Schlatter, P. (2016). On determining characteristic length scales in pressure-gradient turbulent boundary layers. \textit{Phys. Fluids}, 28(5), 055101.

\item Vinuesa, R., Hosseini, S. M., Hanifi, A., Henningson, D. S., \& Schlatter, P. (2017a). Pressure-gradient turbulent boundary layers developing around a wing section. \textit{Flow Turbul. Combust.}, 99, 613--641.

\item Vinuesa, R., \"Orl\"u, R., Vila, C. S., Ianiro, A., Discetti, S., \& Schlatter, P. (2017b). Revisiting history effects in adverse-pressure-gradient turbulent boundary layers. \textit{Flow Turbul. Combust.}, 99(3-4), 565--587.

\item Vinuesa, R., Negi, P. S., Atzori, M., Hanifi, A., Henningson, D. S., \& Schlatter, P. (2018). Turbulent boundary layers around wing sections up to $Re_c= 1,000,000$. \textit{Int. J. Heat Fluid Flow}, 72, 86--99.
\end{References}
\end{document}